\documentclass{eas}
\usepackage{natbib}
\usepackage{graphicx}
\usepackage{amsmath}
\usepackage{amsfonts}
     \usepackage{color}  \definecolor{myred}{rgb}{0.7,0.0,0.2} 
\usepackage{color}  
\definecolor{myred}{rgb}{0.7,0.0,0.2} 
\newcounter{exno}
\newcommand\ednote[1]{\addtocounter{exno}{1}%
\ifodd\value{page}
  \normalmarginpar
  \marginpar[#1]{\raggedright {\sf \color{blue} {\bf (\Roman{exno}}). #1}}
\else
  \reversemarginpar
  \marginpar[#1]{\raggedright {\sf \color{blue} {\bf (\Roman{exno}}). #1}}
\fi}
\newcommand{\corot}{\textsl{CoRoT}}
\def\nropl{374}         % 20090921
\def\nrosistplmult{40}  % 20090921
\newcommand{\frcn}[2]{ {\displaystyle \frac{#1}{#2} } }

\begin{document}

%%-----------------------------
%%      the top matter
%%-----------------------------
\title{Search for Super Earths by Timing of Transits with CoRoT} 
\author{Juan Cabrera}\address{Deutsches Zentrum f{\"u}r Luft- und Raumfahrt, Rutherfordstrasse 2, 12489 Berlin, Germany\\
  LUTH, Observatoire de Paris, CNRS, Universit{\'e} Paris Diderot ; 5 Place Jules Janssen, 92190 Meudon,France} 
\begin{abstract}
We explore the possibility of detecting Super Earths via transit
timing variations with the satellite \corot.
\end{abstract}
\runningtitle{Cabrera: Super Earths by TTV with CoRoT}
\maketitle
%%-----------------------------
%%      your text
%%-----------------------------
\section{Introduction}
\label{sec:intro}

The satellite \corot\, \citep{baglin2006} was launched on 27th December
2006 with a double scientific purpose: the analysis of stellar 
seismology and the detection of extrasolar planets by the method of
transits. So far, 7 transits have been published (see Jean Schneider's
Extrasolar Planets Encyclopaedia\footnote{\rm http://exoplanet.eu}),
but more will come in the near future. Table \ref{tabla:planetas}
gathers the data from these six planets and one brown dwarf. \corot's
photometric precision is below $8\cdot 10^{-4}$ in 2h at $R=15$ \citep{aigrain2009} and candidates are found 
with transit depths of $0.034\%$ \citep{leger2009}. Neptune size planets seem to be
common \citep{gould2006} and, there is something even more interesting: they don't
come alone \citep{mayor2008}. Super Earths, which should also be
numerous, are in the range of detectability of \corot
\citep{leger2009,queloz2009}. In the near future the number of
candidates will increase and we will find ourselves with a collection
of planets whose diversity we can only start to imagine.  

Section \ref{sec:corot} is a short introduction to the
\corot\, mission. Section \ref{sec:ttv} gives a short overview of
different sources of transit timing variations which could allow the
detection of Super Earths with the satellite \corot.

\section{CoRoT}
\label{sec:corot}

\corot\, is an afocal telescope with a 27~cm diameter pupil,
equipped with 4 CCDs ($2048 \times 2048$ pixels each); the pixel scale is 
$2.32''$ and the field of view is $3.05^{\circ}\times2.8^{\circ}.$ The 
selection of observational targets follows two different strategies: the
seismology channel observes a small number (10) of  bright stars
($6<m_v<9$) with a cadence of 32s whereas the exoplanet channel
observes a large number ($\sim 11\,000$) of faint stars ($12<m_v<16$)
every 512s (although a limited number of targets is measured every
32s). \corot\, is placed in a polar low Earth orbit which determines the
observational scheme. The satellite continuously monitors the same
region of the sky during 150 days; but then it has to turn around to
avoid the Sun entering the field of view: these are the \emph{long
  runs}. Immediately before or after the turnaround, the satellite is 
pointed during roughly 20 days to perform a \emph{short run} in a
different direction. Every year, \corot\, observes 2 long runs and 2
short runs providing roughly $40\,000$ light curves. These light
curves are narrowed down in the search for transits and a list of
candidates is built. The most promising candidates are followed up
photometrically and spectroscopically from the ground.  

The photometric follow-up attempts to discover if the transit is on
target or, on the contrary, if it is produced by a background
binary. The PSF of \corot\, is quite large: the flux for each star is
calculated on-board in masks of size 60 pixels on average. In front of
the CCD there is a prism used to produce chromatic light curves with
the aim of distinguishing between stellar (coloured) activity and
(achromatic) transits. However, for faint stars it is not possible
to make this distinction. Large masks raise the probability of
observing background binaries, which are a major source of confusion  
\citep{pont2005}. 

\begin{figure}[t]
  \begin{center}
    \includegraphics[%
      keepaspectratio,%
      totalheight=0.5\textheight,%
      width=\linewidth]{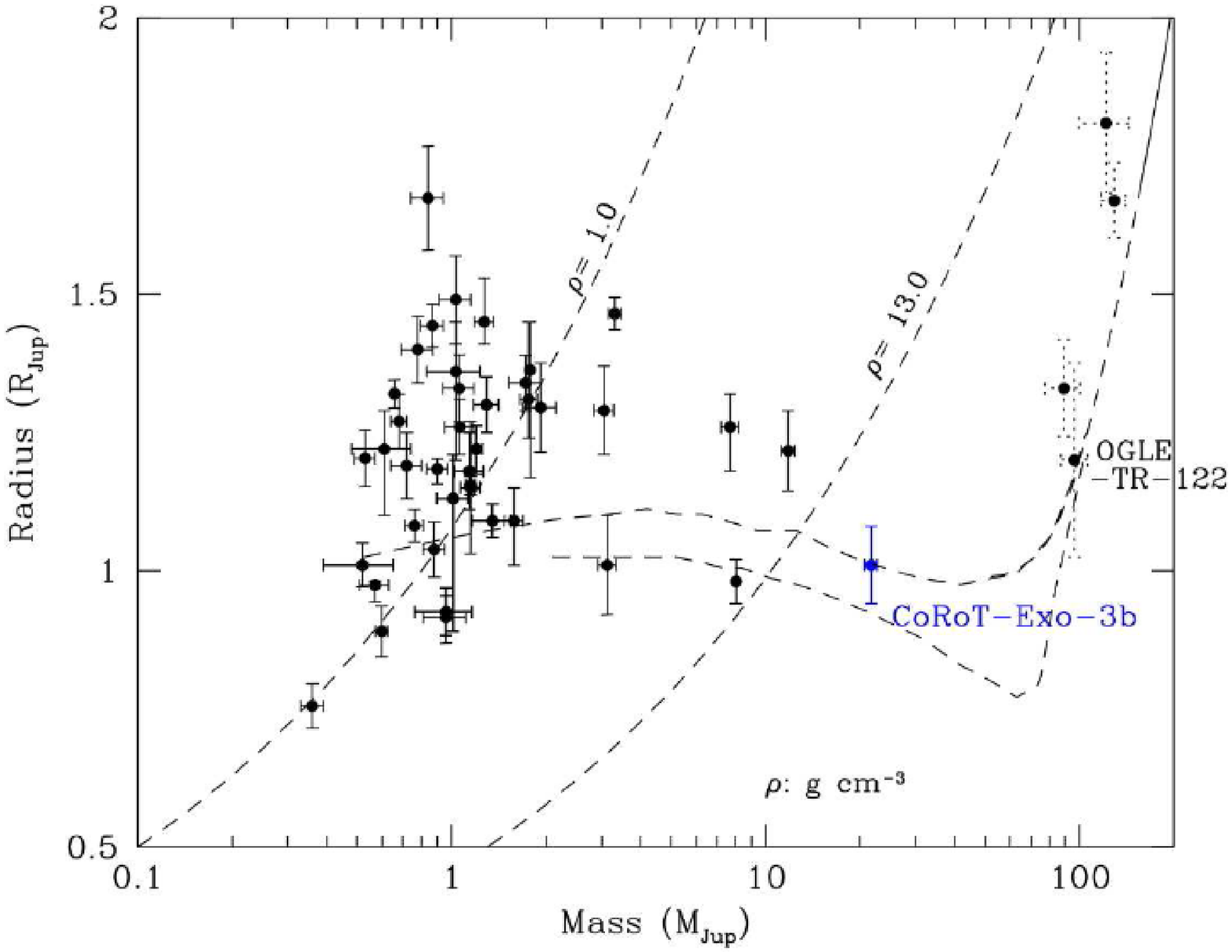}
  \end{center}
  \caption{Figure from \citet{deleuil2008a} showing the mass--radius
  diagram for planets and low mass M stars. CoRoT-3b is
  highlighted. The theoretical isochrones at 10 and 1 Gyr are from
  \citet{baraffe2003}.}
  \label{fig:massradius}
\end{figure}

Spectroscopic follow-up measures the mass of the transiting
object. There is a degeneracy between the mass and the radius of low
mass stellar objects, brown dwarfs and planets (see
Fig.~\ref{fig:massradius}); to confirm the nature of a transiting
object it is mandatory to perform radial velocity measurements and
calculate the object's mass. This can become a bottleneck for the
characterization because the measurement of faint candidates is
challenging.

\citet{deleuil2008b} is a very interesting short summary of
\corot\, and its achievements. Fully detailed recent information about
the technical characteristics of the mission can be found in 
\citet{fridlund2006,barge2008b} and \citet{drummond2008}.  

%\begin{table*}
%  \begin{minipage}[t]{\textwidth}
%    \caption[]{5 transits found by \corot.}
%    \label{tabla:planetas}
%    \centering
%    \renewcommand{\footnoterule}{}
%    \begin{tabular}{*{6}{c}}
%      & exo-1b\footnote{\citet{barge2008a}.} 
%      & exo-2b\footnote{\citet{alonso2008a,bouchy2008,alonso2008c}.}
%      & exo-4b\footnote{\citet{aigrain2008,moutou2008}.}
%      & exo-5b\footnote{\citet{rauer2009}.} 
%      & exo-3b\footnote{\citet{deleuil2008a}.} \\
%      \hline
%      radius & 1.49 & 1.47 & 1.19 & 1.2 & {\it 1.01} \\
%      {\small (Jupiter's radii)} & & & & & \\
%      \hline
%      mass & 1.03 & 3.31 & 0.72 & 0.86 & {\it 21.66} \\
%      {\small (Jupiter's masses)} & & & & &  \\
%      \hline
%      period & 1.51 & 1.74 & 9.2 & 4.0 & {\it 4.25} \\
%      {\small (days)} & & & & & \\
%      \hline
%    \end{tabular}
%  \end{minipage}
%\end{table*}
\begin{table*}
  \begin{minipage}[t]{\textwidth}
    \caption[]{5 transits found by \corot.}
    \label{tabla:planetas}
    \centering
    \renewcommand{\footnoterule}{}
    \begin{tabular}{*{8}{c}}
      & 1b\footnote{\citet{barge2008a}.} 
      & 2b\footnote{\citet{alonso2008a,bouchy2008,alonso2008c}.}
      & 4b\footnote{\citet{aigrain2008,moutou2008}.}
      & 5b\footnote{\citet{rauer2009}.} 
      & 6b\footnote{\citet{fridlund2009}.} 
      & 7b\footnote{\citet{leger2009,queloz2009}.} 
      & 3b\footnote{\citet{deleuil2008a}.} \\
      \hline
      radius & 1.49 & 1.47 & 1.19 & 1.39 & 1.15 & 0.15  & {\it 1.01}  \\
      {\small (Jupiter's radii)}  & & & & & & & \\
      \hline
      mass   & 1.03 & 3.31 & 0.72 & 0.47 & 3.3  & 0.015 & {\it 21.66} \\
      {\small (Jupiter's masses)} & & & & & & & \\
      \hline
      period & 1.51 & 1.74 & 9.20 & 4.04 & 8.89 & 0.85  & {\it 4.25}  \\
      {\small (days)}             & & & & & & & \\
      \hline
    \end{tabular}
  \end{minipage}
\end{table*}

%.....................................................................
\section{Transit Timing Variations}
\label{sec:ttv}

Kepler's laws of motion assign periodical orbits to planets. However,
there are numerous sources of perturbations which produce deviations
from the periodicity. Not only are there differences in the observed minus
calculated (O$-$C) epochs of transits, but also in their durations and
depths. Some possible sources are general relativity effects, the
quadrupolar moment of the gravitational potential of the star, tidal
interaction or even the proper motion of the star
\citep{miralda2002,jordan2008,pal2008,rafikov2008}. But all these 
perturbations act on timescales much longer than the baseline of
\corot\, observations and so, in spite of their interest, are beyond the scope of this study. 

However, there are still several other sources of perturbations acting
on shorter timescales, such us those produced by other planets
\citep{schneider2004,holman2005,agol2005,nesvorny2008}, Trojan planets
\citep{laughlin2002,dvorak2004,ford2007}, moons
\citep{doyle2004,kipping2008b,kipping2009}, orbital eccentricity \citep{kipping2008a}
and the light time effect, the so-called LITE (widely studied in
binary systems, see \citealt{irwin1952,mayer1990,borkovits2003}). In
addition to these, \citet{winn2008} contains a very interesting list 
of the information that can be obtained from transits.

%--------------------------------------
\subsection{Photometric Precision}
\label{subsec:prec}

From \citet{doyle2004}, we can calculate the maximal accuracy of
%%% 3.222222222222 is a precise (but inaccurate) estimate of pi; 
%%% 3.14 is a more accurate (but less precise) estimate of pi --langed
$\delta t_0$ that one can achieve when determining the position of a
transit of length $T_{tr}$ and depth $\Delta L$; the photometric 
accuracy is $\delta L$ and the number of observations is $N$. This
accuracy is:
\begin{equation*}
\delta t_0 = \delta_L \frac{T_{tr}}{2 \Delta L \sqrt{N}}.
\end{equation*}

In \corot, with a photometric accuracy of $0.1\%$, measuring a
transit of depth $1\%$ at the observing cadence of 32s, we can
achieve a timing accuracy on the order of seconds.

%--------------------------------------
\subsection{Multiple Systems}
\label{subsec:sistmult}

On the day this manuscript was submitted, there were \nropl\,extrasolar
planets known, among which most are isolated. But there are already
\nrosistplmult\, known multiple planet systems and in the future, as
Dr. Udry  pointed out in this conference, probably more and more
planets will be found in multiple systems. See also the work by
Dr. Wright in this volume.

We can calculate the perturbations in the time of arrival of the
transit signal of a planet if there is another planet in the system in
an interior orbit. Rigorous calculations are done in
\citet{agol2005}, but we can estimate $\delta t$, the amplitude of
this perturbation, with the expression: 
\begin{equation}
\label{eq:agol}
\delta t = \frcn{P_e}{2 \pi} \frcn{m_i}{m_*+m_i} \frcn{a_i}{a_e};
\end{equation} 
where $P$ stands for period, $m$ for masses, $a$ for the semi-major axis
of the orbits and the subscripts $i$ and $e$ refer to the inner and
outer (exterior) planet respectively. For a Jupiter outer planet
with a period of 20 days around a star of one solar mass, an interior
Super Earth of 11 terrestrial masses would produce a perturbation of 3
seconds, which is within the limits of \corot.

Dynamics in multiple planets systems is a complicated matter 
%(for example see, in this volume, the works by Dr. Laskar, \ednote{Lascar and Laughlin contributions have not been delivered.} Dr. Michtchenko or Dr. Laughlin)
(for example see, in this volume, the work by Dr. Michtchenko) 
 and resonances are one of the most important features
because they enhance the amplitude of these perturbations and could
open the door to the discovery of low mass planets
\citep{holman2005,haghighipour2007}.  

In 2008 alone, at least 8 publications have seen the light on the
detection of this kind of perturbation:
\citet{agol2008,alonso2008c,diaz2008,hrudkov2008,irwin2008,miller2008b,miller2008a,shporer2008}.

%-----------------------
\subsection{Trojan planets}
\label{subsec:troyanos}

In our Solar System, Trojan satellites are a group of asteroids moving
close to the Lagrange points L4 and L5 in 1:1 mean-motion resonance
with Jupiter's orbit. Many efforts have been done in the search for
these kind of objects in extrasolar systems (see for example
\citealt{moldovan2008} and \citealt{madhusudhan2008}). Bodies in these
orbits are stable \citep{ford2006,dvorak2004} and can be found
not only photometrically or by radial velocity, but also by the
timing variations that they produce in the transits of the planet
whose orbit they share. We can estimate the amplitude of this
perturbation: 
\begin{equation}
\label{eq:troyano}
\delta t = \frac{M_{Trojan}}{M_{planet}} \frac{ {\alpha} }{2 \pi} P_{planet};
\end{equation}
for a Trojan object with the mass of the Moon and a transiting planet
with the mass of Jupiter in a 20 day orbit, 
$\alpha$ being the typical angle
involved in the calculation, with $\alpha \sim 30$ degrees (see the references given above 
for justification), the amplitude of the perturbation is about 5 
seconds. Needless to say, if the transiting planet is a Super Earth, 
this perturbation is far more important. Another speculative hypothesis
is the existence of massive Trojan planets. If the ratio between the
mass of the transiting planet plus the mass of the Trojan over the
mass of the star is below $\sim 1/27,$ the system can be stable; this
opens the possibility of Trojan Super Earths (see
\citealt{nauenberg2002,schwarz2007} and references therein). 

%------------------------------------
\subsection{LITE}
\label{subsec:c4_roemer}

LITE was first used by the astronomer Ole R{\"o}mer, working in Paris
Observatory with Jean-Dominique Cassini, to measure the speed of light
\citep{roemer1676}. Nowadays it is used to find hidden companions to
binary systems, even those of planetary mass
\citep{deeg2008,lee2008}. But we can find the same effect in multiple
planet systems \citep{schneider2005}. The reflex motion induced in
our Sun by Jupiter has an amplitude of one solar radius, which light
covers in 2 seconds. If we observe the transits of an inner planet and 
there is an outer planet in the system, the amplitude of the LITE 
perturbation is: 
\begin{equation}
\label{eq:c4_roemer}
\delta t = 2 \frac{m_e}{M_*} \frac{a_e}{c};
\end{equation} 
which is of the order of 0.1 seconds for the time baselines of
\corot\, (and in this case, we must concede that this favors
the detection of high mass companions and not of Super Earths).

%--------------------------
\subsection{Moons}
\label{subsec:c4_satelites}

No moon has yet been detected around any extrasolar planet, although
their existence is expected \citep{sartoretti1999}. However, their 
detection is difficult \citep{brown2001}. 
In general, it
is not an easy task to
estimate the magnitude of the perturbation because it depends on the
orbit of the satellite around the planet; and 
for planets within the specific period range detectable by \corot,
we don't yet have any
clue as to how much this estimate may be.
Nevertheless, reasonable assumptions in the general case give 
perturbations under 1~s, which is below \corot\, limits.

However, we point out the possibility of finding binary planets
\citep{cabrera2007}. Binarity is common among stars, from bright
massive objects down to brown dwarfs; and we can also find binary
objects from the size of trans-neptunian objects down to
asteroids. Binarity should be possible among planets and those
systems will produce peculiar transit signals. 

% \ednote{The paper would benefit from at least short conclusions.}
%.....................................................................
\section{Conclusions}
\label{sec:conclusion}

The photometric precision achieved by \corot\,allows the detection of
Super Earth planets in transit; but here we have shown that also
non-transiting Super Earths could be detected in multiple systems by
the perturbations they might produce in transiting planets. We have
shown several possible scenarios and discussed their limitations. 
 
%%-----------------------------
%%      bibliography
%%-----------------------------
\bibliographystyle{astron}
\bibliography{bibl}
\end{document}